\documentclass[10pt,a4paper]{article}

\usepackage{cogsci}
\cogscifinalcopy 

\usepackage{pslatex}
\usepackage{float} 
\usepackage{graphicx}
\usepackage{setspace}
\usepackage{hyperref}
\usepackage{apacite}

\title{Mapping fMRI Signal and Image Stimuli in an Artificial Neural Network Latent Space: Bringing Artificial and Natural Minds Together\\[0.4cm]}

\author{
  \begin{tabular}{cc}
    {\large \bf Cesare Maria Dalbagno\textsuperscript{1}} & {\large \bf Manuel de Castro Ribeiro Jardim\textsuperscript{1}} \\
    {\normalfont\texttt{c.m.dalbagno@tilburguniversity.edu}} & {\normalfont\texttt{m.decastroribeirojardim@tilburguniversity.edu}} \\
  \end{tabular} \\
  \\
  {\large \bf Mihnea Angheluță\textsuperscript{1}} \\
  {\normalfont\texttt{m.aangheluta@tilburguniversity.edu}}\\ [0.3cm]
  \textsuperscript{1}Department of Cognitive Science and Artificial Intelligence, Tilburg University
}

\begin{document}

\maketitle

\begin{abstract}
The goal of this study is to investigate whether latent space representations of visual stimuli and fMRI data share common information. Decoding and reconstructing stimuli from fMRI data remains a challenge in AI and neuroscience, with significant implications for understanding neural representations and improving the interpretability of Artificial Neural Networks (ANNs). In this preliminary study, we investigate the feasibility of such reconstruction by examining the similarity between the latent spaces of one autoencoder (AE) and one vision transformer (ViT) trained on fMRI and image data, respectively. Using representational similarity analysis (RSA), we found that the latent spaces of the two domains appear different. However, these initial findings are inconclusive, and further research is needed to explore this relationship more thoroughly.

\end{abstract}

\section{Introduction}

Our initial and ambitious goal was to reconstruct fMRI data from image stimuli and vice versa. To achieve this, we planned to train two autoencoders (AEs: \cite{huang_modeling_2018}): one on image stimuli and the other on fMRI data collected from subjects viewing those stimuli. We aimed to transform one latent space into the other, enabling model stitching \cite{moschella_relative_2023}. The envisioned architecture would combine the encoder from one model, the transformed latent space, and the decoder from the other model, allowing the reconstruction of one type of data from the other. However, due to time constraints and limited computational resources, this could not be achieved. Nonetheless, the results presented provide a necessary first step to assess the feasibility of the initial objective.

Understanding the neural representation of visual stimuli is a fundamental question in neuroscience. The ability to decode and reconstruct visual stimuli from brain data, particularly fMRI, has been explored extensively \cite{du_fmri_2022,kay_identifying_2008,lin_dcnn-gan_2019}. Recent advances in machine learning have introduced powerful tools for this purpose, including autoencoder models and vision transformers. Autoencoder models excel in learning representations due to their ability to reduce dimensionality while preserving essential features of the input data. This is crucial for effective feature extraction and understanding complex data \cite{guo_deep_2017}. Vision transformers have become a research hotspot for image feature extraction due to their ability to handle high-dimensional data and capture complex patterns through attention mechanisms \cite{zheng_lightweight_2024}.

Recent research has shown that well-performing neural networks often learn similar representations across various architectures, tasks, and domains \cite{li_convergent_2016,morcos_insights_2018}. Representational similarity analysis is a method used to compare the similarity of representations in the models. RSA does not directly map brain activity to specific measures but instead assesses the similarities within each NN's representational space by examining their similarity matrices \cite{kriegeskorte_representational_2008}. This approach allowed us to examine different types of data whose embeddings could not be directly compared. Stimuli were embedded in a high-dimensional feature space, and pairwise similarities were calculated.

\section{Methods}

\subsection{Preprocessing}
We used the BOLD5000 dataset \cite{chang_bold5000_2019}, a large-scale fMRI study of visual scene processing, for our analysis. This dataset includes data from 4 participants and 4,916 unique scenes collected over 16 sessions, with images from Scene Images, COCO, and ImageNet datasets \cite{lin_microsoft_2014,russakovsky_imagenet_2015,xiao_sun_2010}. Participants viewed images for 1 second with 9 seconds of blank screen fixation, and their responses and physiological data were recorded. We pruned the dataset to only contain the “.nii.gz” and the “.tsv” files from each run of each session for ease of processing. We also modified the name of each run accordingly. 

We employed the nilearn and nibabel libraries to handle and visualize neuroimaging data \cite{abraham_machine_2014}. The fMRI images were preprocessed using the following steps. First, fMRI data was loaded using the nibabel library. Activation images were generated and saved for each time step in the fMRI data using nilearn’s plotting functions. The images were subsequently reshaped to 672x672 pixels and converted to tensors. After the transformations, the tensors were loaded into the autoencoder for further processing. All the code used in this study is available on GitHub for reproducibility \cite{jefftheninja57_jefftheninja57research_workshop_2024}.

\subsection{Autoencoder Training}
To learn meaningful representations of the brain data we implemented a convolutional autoencoder using PyTorch \cite{paszke_pytorch_2019}. The encoder is composed of 5 convolutional layers (kernel = 3, stride = 2, padding = 1, activation function = ReLU) reducing the input image of shape 3x672x672 to an output of 256x21x21. The decoder performs the image reconstruction through 5 layers of transposed convolutions with the same parameters as the encoder and a sigmoid activation function in the last layer (Figure 1).

\begin{figure}[H]
\begin{center}
\includegraphics[width=8cm]{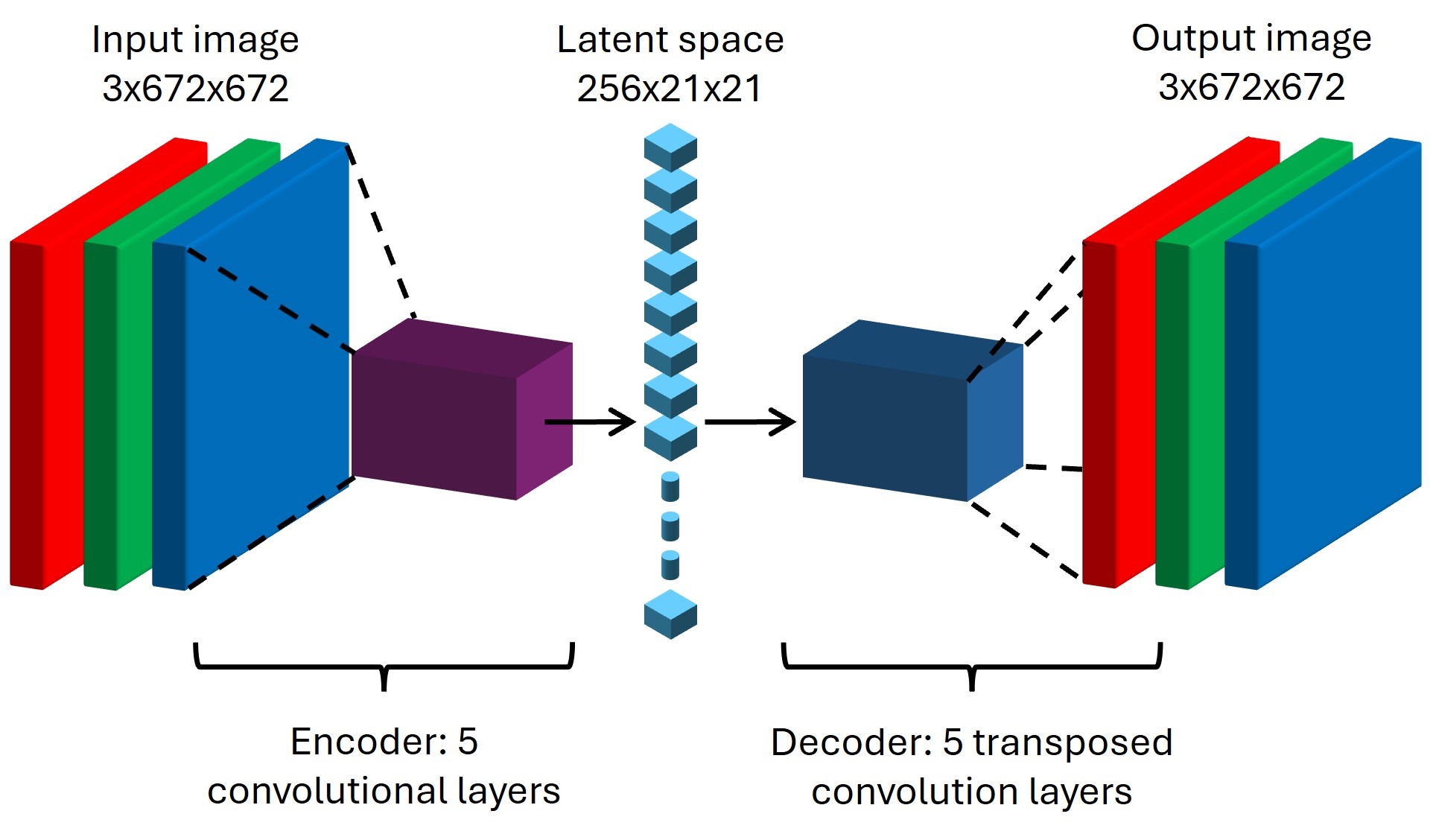}
\end{center}
\caption{Convolutional autoencoder} 
\label{Figure 1}
\end{figure}

We split the data subject-wise to build the training and test set. Subjects 1 to 3 were used for training, and subject 4 for testing, to make sure the model generalized well. We implemented early stopping and experimented with adding more layers and dropouts to avoid overfitting. Overall, after 100 epochs, we achieved a training loss of 0.001 and a test loss of 0.002, calculated as the mean squared error (MSE) on the reconstructed image (Figure 2).

\begin{figure}[H]
\begin{center}
\includegraphics[width=8cm]{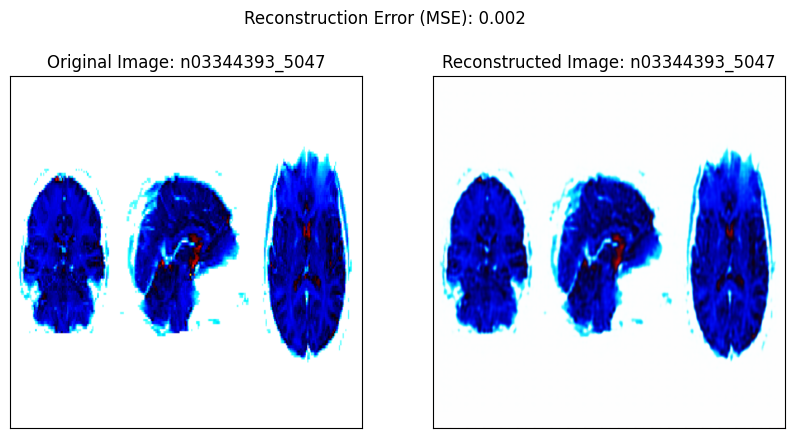}
\end{center}
\caption{Original vs reconstructed images} 
\label{Figure 2}
\end{figure}

\subsection{Embedding Extraction and RSA Analysis}
Next, we randomly split the stimuli images into 4 sets and retrieved the embeddings of those 4 sets subject-wise, without resampling, ensuring an even distribution of the learned representations. We used a pre-trained ViT to compute the embeddings of the image stimuli, specifically Google’s vit-base-patch16-224-in21k \cite{noauthor_googlevit-base-patch16-224-in21k_nodate}. To get the embeddings, we loaded the model and passed our image stimuli through it, using a function to extract the embeddings for each image.

\begin{figure}[H]
\begin{center}
\includegraphics[width=8cm]{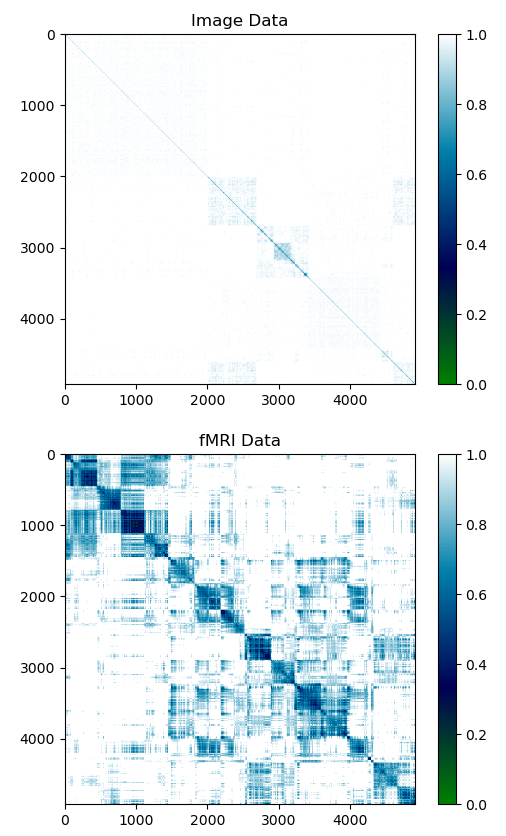}
\caption{Two correlation matrices for image data (top) and fMRI (bottom), RSA value r = 1.672e-02, $(p<0.001)$}
\end{center}
\label{Figure 3}
\end{figure}

Once the embeddings of the image stimuli and its matching fMRI data have been acquired and set in the same order, we went on to reduce the dimensionality of the data. This is done as a precursor to performing RSA, and is especially important on the data gathered from the AE. Due to the similar nature of the images, and the number of dimensions of their embeddings (256), all the embeddings were highly correlated. To correct this, we standardized all the embeddings and performed dimensionality reduction using Truncated singular value decomposition (SVD) from the Scikit-learn package \cite{noauthor_truncatedsvd_nodate}. We chose TruncatedSVD due to its ability to find non-linear relationships in the data. Once reduced, the fMRI data had 38 dimensions instead of the original 256. The same process was followed for the image stimuli embeddings due to computational constraints, as the original embeddings from the ViT have 151,296 dimensions. Once the dimensions were reduced by Truncated SVD, we used 715 dimensions that explain around 70 \% of the variance.

We then proceeded to calculate the dissimilarity matrices as this is the first step in conducting RSA. A dissimilarity matrix is a normal correlation matrix where each correlation coefficient is subtracted from 1, inverting the interpretation of the matrix. This inverts the interpretation of the matrix, high values show high dissimilarity and low correlation, while low values show low dissimilarity and high correlation. All correlations were  calculated using Pearson's correlation. Using the two dissimilarity matrices each representing the degree of correlation of the latent space of the data, we proceeded to correlate the two matrices. This was also done using Pearson's correlation and is the final step in RSA, from which we obtain a correlation value (Pearson's r) and a confidence value (Figure 3).

\section{Results}

The results we observe from performing RSA on the two separate latent spaces conclude that the two are not correlated, r(4999) = 1.672e-02, $(p<0.001)$. There are several potential reasons to deem this outcome not fully satisfying and many possible improvements.

The ViT we used was trained on ImageNet, which may have led to a more meaningful understanding of the ImageNet images compared to those from the COCO and Scene datasets. This discrepancy could lead to a difference in correlation of embeddings and limit our ability to generalize. To address this, we could fine-tune the ViT with our data or construct another (AE) to generate the image embeddings.

We were also limited by our knowledge and experience when working with fMRI data. The representation we chose to train the AE might not be the most suitable for fMRI data, and could therefore decrease the meaning of our analysis.

While our hypothesis proved untrue, we successfully built and trained the AE. Despite using unorthodox methods, we trained the AE effectively, and it demonstrated an understanding of the data, as evidenced by the fMRI data correlation matrix. With a low reconstruction error, it proved to be a reliable method for reconstructing the images we fed into it, even though it may not have been the best representation for our fMRI data analysis.

\section{Discussion and Conclusion}

By comparing embeddings from the AE and the ViTs using RSA, we found no significant correlation between the latent space representations of visual stimuli and fMRI data, suggesting that current models may not effectively capture shared information between these data types. Despite inconclusive results, this study reveals interesting insights as well as notable limitations and future research directions. Overall, these findings align with the complexity of the task, as demonstrated by low benchmarks established in previous studies. 

In assessing the results, it appears that the ViT's pre-training on ImageNet may have introduced biases, contributing to the lack of alignment between the two latent spaces. Additionally, limitations in our preprocessing methods and AE model architecture, as well as limited computational resources and time constraints, may have restricted our approach.   
Future research should include fine-tuning models on target datasets, exploring different preprocessing techniques, and developing more sophisticated model architectures. Additionally, aligning the latent spaces through geometric manipulations to ensure they capture semantically meaningful distributions could improve results.  Exploring methods for semantic alignment \cite{maiorca_latent_2023} could facilitate the retrieval of images from fMRI data and vice versa, potentially through model stitching. Addressing these gaps could advance our understanding of neural representations and increase the interpretability of artificial neural networks.

\section{Statement of Technology}
GitHub Copilot was used to assist us with refining the code and automating repetitive tasks. 
GPT 4o helped us rewrite and rephrase some sentences while adjusting the text to LateX format. 
Other online tools for spelling errors correction (such as Grammarly) have also been used individually to aid us in the writing phase.

\bibliographystyle{apacite}

\setlength{\bibleftmargin}{.125in}
\setlength{\bibindent}{-\bibleftmargin}

\bibliography{rw_paper.bib}

\end{document}